\documentclass[5p]{elsarticle}
\usepackage{hyperref}
\usepackage[T1]{fontenc} 
\usepackage{graphicx}
\usepackage{float}
\usepackage{adjustbox}
\usepackage{natbib}
\usepackage{hyperref}
\usepackage{amsmath}
\usepackage{amsthm}
\usepackage{mathtools, cuted}
\usepackage{adjustbox}
\usepackage{upgreek}
\usepackage{xcolor}
\usepackage{textcomp}
\usepackage{makecell}
\journal{~}

\biboptions{authoryear}

\newsavebox\MBox

\begin{document}

\begin{frontmatter}

\title{Increased dose rate precision in combined $\alpha$ and $\beta$ counting in the $\mu$Dose system - a probabilistic approach to data analysis}

\author[SUT]{Konrad Tudyka}
\ead{konrad.tudyka@polsl.pl}

\author[SUT]{Andrzej Bluszcz}

\author[SUT]{Grzegorz Por\k{e}ba}

\author[SUT]{Sebastian Mi\l{}osz}

\author[SUT]{Grzegorz Adamiec}

\author[UDOSE]{Aleksander Kolarczyk}

\author[GI]{Thomas Kolb}

\author[GI]{Johanna Lomax}

\author[GI]{Markus Fuchs}

\address[SUT]{Silesian University of Technology, Institute of Physics - Centre for Science and Education, Division of Radioisotopes, ul. S. Konarskiego 22B, 44-100 Gliwice, Poland}
\address[UDOSE]{miDose Solutions, ul. Wolno\'sci 234b/4, 41-800 Zabrze, Poland}
\address[GI]{Justus-Liebig-University Giessen, Department of Geography, 35390 Giessen, Germany}

\begin{abstract}
The $\mu$Dose system was developed to allow the measurement of environmental levels of natural radioactive isotopes.  The system records $\alpha$ and $\beta$ particles along with four decay pairs arising from subsequent decays of $^{214}$Bi/$^{214}$Po, $^{220}$Rn/$^{216}$Po, $^{212}$Bi/$^{212}$Po and $^{219}$Rn/$^{215}$Po. Under the assumption of secular equilibrium this allows to assess the specific radioactivities of $^{238}$U, $^{235}$U, $^{232}$Th decay chains and $^{40}$K. This assessment provides results with uncertainties which are correlated and, thus, require the development of an error estimation methodology which considers this issue. Here we present two different approaches for uncertainty propagation based on Monte Carlo and Bayesian methods. Both approaches produce statistically indistinguishable results and allow significantly better dose rate precision than when the correlations are not accounted for. In the given example, the dose rate precision is improved by a factor of two.

\end{abstract}


\end{frontmatter}


\section{Introduction}

In trapped charge dating, the age is determined from the equivalent of the total absorbed radiation dose and the radiation dose rate. The radiation dose rate is often measured through the detection of ionising radiation or derived from other methods that allow us to assess the specific activities of radioactive elements that contribute to the dose rate in the natural environment. In many cases the major dose rate contributors are $^{238}$U, $^{235}$U, $^{232}$Th decay chains and the $^{40}$K isotope.\\
\indent The dose rate calculation for trapped charge dating requires a number of corrections, which have been implemented in several tools routinely used in this dating method \citep{Grun2009,Durcan2015}. In addition, other programs were developed but not formally published, for example ADELE by \cite{Kulig2005} or Dose4Win by \cite{Bluszcz2001}. Such programs offer a convenient way of obtaining the dose rate which can be directly used in trapped charge dating.\\
\indent The $\mu$Dose system \citep{Tudyka2018a} contains an in-built module for obtaining corrected dose rates. In contrast to the above mentioned programs, this module handles correlated uncertainties that arise from the performed measurements of the $\alpha$ and $\beta$ count rates, as well as from four different decay pairs (\textit{ibid.}). The dose rate calculation procedure is fully controlled by an intuitive graphical user interface making the process convenient and straightforward.
In this current work we describe how correlated uncertainties are accounted for in the dose rate calculation algorithms which are implemented in the $\mu$Dose system. Such an approach can be developed for any system characterised by correlated uncertainties. 
The error propagation can be achieved through either a Monte Carlo method or a probability distribution propagation (based on the Bayes theorem). Here we develop and compare both approaches. Dose rate adjustment for alpha efficiency, grain size distribution, chemical etching, gamma scaling and water content, which are all used in trapped charge dating, are also considered.\\
\indent Finally, we show how the inclusion of correlated uncertainties allows the precise determination of the dose rate when compared to uncorrelated uncertainties.

\section{Dose rate calculation module}
\begin{figure*}
\includegraphics[width=160mm]{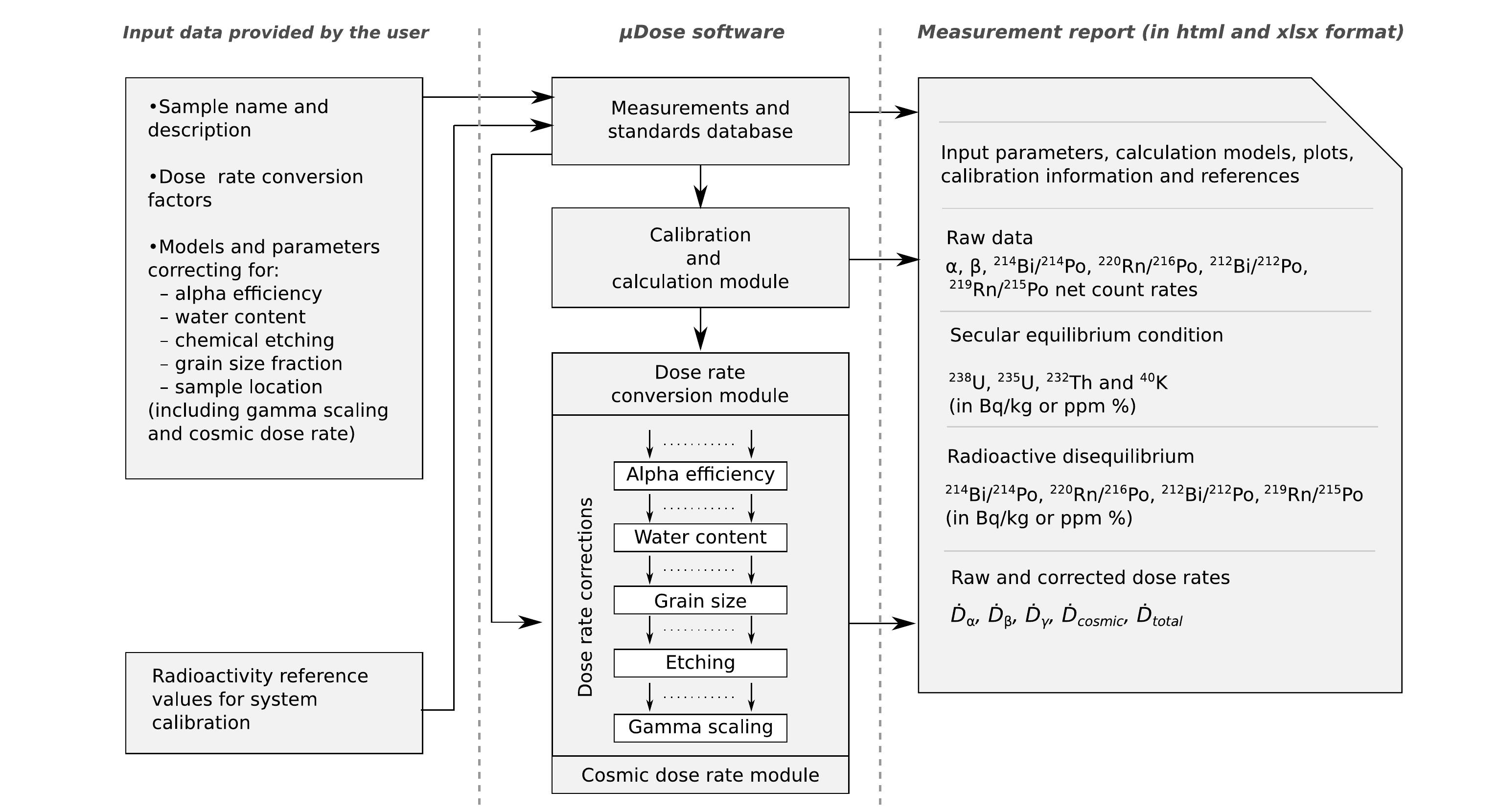}
\caption{\label{Fig1}Dose rate calculation organisation in the $\mu$Dose system.}
\end{figure*}

In trapped charge dating methods, the annual dose is typically determined under the infinite matrix assumption \citep[e.g.][]{Guerin2012} or with the use of Monte Carlo simulations of charge transport  \citep[e.g.][]{Martin2014} when the concentrations of radioisotopes and geometry are given. \\
\indent The dose rate calculation module is organized as shown in the block diagram given as Fig. \ref{Fig1}. First, the activities (or concentrations) of $^{238}$U, $^{235}$U, $^{232}$Th decay chains and their daughters, as well as $^{40}$K isotope are calculated as described in the following section. Next, dose rate distributions are obtained through the application of either a Monte Carlo or a Bayesian approach. Finally, factors like water content, grain size and other effects that are commonly used in trapped charge dating are accounted for. This is briefly described in section \ref{Dose rate conversion and additional corrections}. The module returns the results in the form of reports giving the $\alpha$, $\beta$, $\gamma$ and the total dose rate distributions, with theirs respective means and uncertainties. These values can be directly used for trapped charge dating.

\subsection{Decay chains and $^{40}$K activity assessment}
$\mu$Dose is equipped with a pulse analyzer \citep{Miosz2017} that allows detection of $\alpha$ and $\beta$ particles along with four types of decay pairs. A detailed description on how the specific activities of $^{238}$U, $^{235}$U, $^{232}$Th decay chains and  $^{40}$K are assessed is given by \cite{Tudyka2018a}. Before presenting this novel approach, the mathematical procedure can be summarised in the following steps:
\begin{enumerate}
\item First, a secular equilibrium in the radioactive chains and an absence of other radioactive isotopes is assumed.
\item Next, the solutions are constrained to yield the, so-called, natural uranium (U-nat) that has a known $^{238}$U/$^{235}$U isotope ratio \citep[e.g.][]{Brennecka2018,Uvarova2014}
\begin{equation}
\frac{^{238}U}{^{235}U}=\frac{a_{U-238}/\lambda_{U-238}}{a_{U-235}/\lambda_{U-235}}=137.88.\label{eq1}
\end{equation}

\item Finally, we assume, like others \citep[e.g.][]{Aitken1985}, that $^{40}$K and $^{87}$Rb are found together in the natural environment. For example \cite{Warren1978} gives an average $^{87}$Rb level of 50~ppm of natural rubidium per 1 \% of potassium.
\end{enumerate}

Using the above assumptions, the following system of equations is formulated:

\begin{strip}
\begin{equation}
\begin{aligned}
&r_{\alpha} = k_{\alpha, Th-232}a_{Th-232}+k_{\alpha, U-238}a_{U-238}+k_{\alpha, U-235}a_{U-235}\\ \label{eq2}
&           = k_{\alpha, Th-232}a_{Th-232}+(k_{\alpha, U-238}+k_{\alpha, U-235}\tfrac{\lambda_{U-235}}{137.88\lambda_{U-238}})a_{U-238},\\
&r_{\beta}  = k_{\beta, Th-232}a_{Th-232}+k_{\beta, U-238}a_{U-238}+k_{\beta, U-235}a_{U-235}+k_{\beta, K-40}a_{K-40},\\
&           = k_{\beta, Th-232}a_{Th-232}+(k_{\beta, U-238}+k_{\beta, U-235}\tfrac{\lambda_{U-235}}{137.88\lambda_{U-238}})a_{U-238}+k_{\beta, K-40}a_{K-40},\\
&r_{Bi-212/Po-212}  =k_{Bi-212/Po-212}a_{Th-232},\\
&r_{Bi-214/Po-214}  =k_{Bi-214/Po-214}a_{U-238},\\
&r_{Rn-220/Po-216}  =k_{Rn-220/Po-216}a_{Th-232},\\
&r_{Rn-219/Po-215}  =k_{Rn-219/Po-215}a_{U-235}= k_{Rn-219/Po-215}\tfrac{\lambda_{U-235}}{137.88\lambda_{U-238}}a_{U-238}.
\end{aligned}
\end{equation}
\end{strip}
Here $r$ represents the net count rates of the detected events indicated in subscripts, $k$ is the calibration coefficient for the given system, the decay pairs are indicated in subscripts, and $a$ refers to specific activities of the isotopes indicated in subscripts. 

Eqs. \ref{eq2} can be rewritten in matrix form:
\begin{strip}
\begin{equation}
\label{eq3}
\begin{bmatrix} r_{\alpha}\\ r_{\beta}\\ r_{Bi-212/Po-212}\\ r_{Bi-214/Po-214}\\ r_{Rn-220/Po-216}\\ r_{Rn-219/Po-215} \end{bmatrix}
=
\begin{bmatrix}
  k_{\alpha, U-238}+k_{\alpha, U-235} \frac{\lambda_{U-235}}{137.88\lambda_{U-238}}  & k_{\alpha, Th-232}& 0  \\
  k_{\beta, U-238} +k_{\beta, U-235} \frac{\lambda_{U-235}}{137.88\lambda_{U-238}}  & k_{\beta, Th-232}&  k_{\beta, K-40} \\
  0 & k_{Bi-212/Po-212} & 0 \\
  k_{Bi-214/Po-214}& 0 & 0 \\
  0 & k_{Rn-220/Po-216} & 0 \\
  k_{Rn-219/Po-215} \frac{\lambda_{U-235}}{137.88\lambda_{U-238}}& 0 & 0
\end{bmatrix}
\begin{bmatrix}a_{U-238}\\ a_{Th-232}\\ a_{K-40}\end{bmatrix} 
\end{equation}
\end{strip}
or more concisely

\begin{equation}
\mathbf{r} = \mathbf{k}\mathbf{a}.\label{eq4}
\end{equation}

\subsection{Monte Carlo module}\label{Monte Carlo module}
$\mu$Dose software solves  $\mathbf{a}$ in Eq. \ref{eq4} by means of weighted least squares method yielding an estimate of the activities, $\boldsymbol{\upmu}$, and a corresponding covariance matrix $\boldsymbol{\Upsigma}$. The matrices have the forms where $\boldsymbol{\upmu} = [\mu_{U-238}, \mu_{Th-232}, \mu_{K-40}]^\mathrm{T}$  and
\begin{equation}
\boldsymbol{\Upsigma}=
\begin{bmatrix}
\sigma_{U-238}^2 & \sigma_{U-238, Th-232}& \sigma_{U-238, K-40} \\
\sigma_{Th-232, U-238} & \sigma_{Th-232}^2 & \sigma_{Th-232, K-40}\\
\sigma_{K-40, U-238} & \sigma_{K-40, Th-232} & \sigma_{K-40}^2
\end{bmatrix}.
\end{equation}

All non-diagonal entries in the $\boldsymbol{\Upsigma}$ matrix are typically negative since $^{238}$U, $^{232}$Th and $^{40}$K are negatively correlated. This is used in the Monte Carlo module where $\boldsymbol{\upmu}$ and $\boldsymbol{\Upsigma}$ are used do draw $N$ random values from the multivariate normal distribution \citep{Duda2001}. This gives $N$ matrices $\mathbf{a}_i = [a_{i,U-238}, a_{i,Th-232}, a_{i,K-40}]^\mathrm{T}$ which contain $^{238}$U, $^{232}$Th and $^{40}$K specific activities. Such randomly selected sets of $^{238}$U, $^{232}$Th and $^{40}$K values form statistically plausible solutions of Eq. \ref{eq4} for given count rates, $\mathbf{r}$, and their uncertainties, $\mathbf{u}(\mathbf{r})$.
Further processing of random $\mathbf{a}_i$ values is described in detail in the section \ref{Dose rate conversion and additional corrections}.

\subsection{Probabilistic module}\label{Probabilistic module}
As an alternative approach to the Monte Carlo module, we can use a probabilistic approach where, instead of generating a population of solutions, we generate a discrete probability density function for solutions of Eq. \ref{eq4}. To do this, we create a set of matrices
\begin{align*}
\mathbf{a}_i = [a_{i,U-238}, a_{i,Th-232}, a_{i,K-40}]^\mathrm{T}
\end{align*}
of statistically plausible solutions, that are evenly spaced. Subsequently, using Eq. \ref{eq4} we map matrices $\mathbf{a}_i$ to $\mathbf{r}_{i,map}$ by calculating $\mathbf{r}_{i,map}=\mathbf{k}\mathbf{a}_i$. the mapped  $\mathbf{r}_{i,map}$ experimental data $\mathbf{r}$ and their uncertainties $\mathbf{u}(\mathbf{r})$ are then used to assign likelihoods $P_{l,i}$ for each mapped $\mathbf{a}_i$. This is completed by determining:

\begin{equation}
P_{l,i} = \prod_{j=1}^{6} f(r_{i,map,j}|r_j, u(r_j)).\label{eq5}
\end{equation}

Here, $f$ is the probability density of the normal distribution with a mean value $r_j$, standard deviation $u(r_j)$, $j$ indicates rows of matrices $\mathbf{r}_{i,map}$, $\mathbf{r}$ and $\mathbf{u}(\mathbf{r})$, i.e. iteration over the $\alpha$, $\beta$, $^{212}$Bi/$^{212}$Po, $^{214}$Bi/$^{214}$Po $^{220}$Rn/$^{216}$Po and $^{219}$Rn/$^{215}$Po count rates respectively.

If $^{238}$U, $^{232}$Th or $^{40}$K specific activity values and uncertainties are known from a different measurement (e.g. HPGe) we can assign that prior distribution i.e. $P_{r,i}$ values assuming a normal distribution. The dissimilarity of two datasets i.e. $P_{l}$ and $P_{r}$ is then measured by the Bhattacharyya distance \citep{Bhattacharyya1943}. If no U, Th or K contents are known then all prior probabilities $P_{r,i}$ are set to the same value, e.g. 1.
Mapped posterior distribution i.e. $P_{o,i}$ for each $\mathbf{a}_i$ matrix can be calculated as:
\begin{equation}
P_{o,i}  \propto P_{l,i} P_{r,i}. \label{eq6}
\end{equation}
Finally, $P_{o,i}$ values mapped on $\mathbf{a}_i$ matrices are forming discrete probability density function (PDFs) that can be used to asses $^{238}$U, $^{232}$Th or $^{40}$K content.

\subsection{Dose rate conversion and additional corrections}\label{Dose rate conversion and additional corrections}

Dose rate conversion factors had been reported frequently in the literature \citep[e.g.][]{AA1988,Guerin2011,Liritzis2012,Cresswell2018}. In this module, the user can choose which dose rate conversion factors should be utilised in the calculations. As default, the newest values summarized by \cite{Cresswell2018} with their uncertainties are applied.\\
\indent To obtain dose rates in the Monte Carlo module, all drawn $\mathbf{a}_i$ matrices from the multivariate normal distribution are converted to $\alpha$, $\beta$ and $\gamma$ dose rates. Next, the means with standard deviations of those values are calculated.\\
\indent In the probabilistic module, data $\mathbf{a}_i$ matrices are converted to $\alpha$, $\beta$ and $\gamma$ dose rates. Next, values $P_{s,i}$ are assigned to each $\mathbf{a}_i$ used as the weighted means and standard deviations can be determined.\\
\indent In trapped charge dating, a number of different adjustments are applied to obtain the proper annual dose rate. These adjustments have been summarized and thoroughly discussed previously \citep[e.g.][]{Durcan2015}. Therefore, we will concentrate on a brief overview of the options currently available in the $\mu$Dose-system.
\begin{enumerate}[(a)]
\item The $a$-value correction is described by \cite{AITKEN1975}, \cite{Aitken1985a} and \cite{Aitken1985}. The user can set the $a$-value and its estimated uncertainty according to the most recent and reliable reports \citep[e.g.][]{Kreutzer2018,Kreutzer2014,Mauz2006,Rees-Jones1995,Schmidt2018}. Here we assume a normal distribution for the uncertainty of the $a$-value.
\item Grain size correction can be performed for $\alpha$ radiation as given by \cite{Bell1980}, \cite{Brennan1991}, and \cite{Fleming1979}. For the $\beta$ radiation the user can use data provided by  \cite{Fleming1979}, \cite{Mejdahl1979}, \cite{Brennan2003} or \cite{Guerin2012} data. In addition, \cite{Guerin2012} also provides correction values for feldspar. In this correction, we assume an uniform distribution within the given range of grain sizes.
\item Chemical HF etching is frequently used to remove the  $\alpha$ contribution in grains. This effect can be accounted for optionally by: a) \cite{Fleming1979} for either etching time in minutes or depth in $\mu$m for $\alpha$\textquotesingle s and $\beta$\textquotesingle s , b) \cite{Bell1979} for  $\alpha$\textquotesingle s and $\beta$\textquotesingle s for etched depth in $\mu$m, c) \cite{Bell1979} data for  $\alpha$\textquotesingle s and \cite{Brennan2003} data for $\beta$\textquotesingle s for etched depth in $\mu$m.
\item Water content correction is calculated according to \cite{Aitken1985} or with further modifications \citep[e.g.][]{Aitken1990a}. The user can set an estimated saturation content and fraction of saturation. Moreover, the user may choose the distribution from normal, uniform or various triangular distributions which will affect the propagation of uncertainty.
\item Gamma dose rate scaling for a given depth is performed as given by \cite{Aitken1985}. This is completed to account for a lower gamma dose at shallow depths up to  ca. 30~cm. In addition, the user can scale for a given density and its estimated uncertainty with an assumed normal distribution.
\item For user convenience, $\mu$Dose can also calculate cosmic dose rates for given a latitude, longitude, altitude, depth and a given sediment overburden density as well as its uncertainty.
	
\end{enumerate}

The uncertainties in both Monte Carlo and Probabilistic modules of the final $\alpha$, $\beta$, $\gamma$ and cosmic dose rates are calculated using a dedicated Monte Carlo simulation.

\section{Results and discussion}

To compare both modules we used IAEA-RGU-1, IAEA-RGTh-1, and IAEA-RGK-1 standards from the International Atomic Energy Agency \citep{IAEA}. The IAEA-RGU-1 and IAEA-RGTh-1 are produced using uranium and thorium ores that are mixed with floated silica powder. Decay chains present in those reference materials are considered to be in secular equilibrium with the parent radioactivity. The IAEA-RGK-1 reference material is produced using high purity (99.8\%) potassium sulfate. IAEA-RGU-1, IAEA-RGTh-1, and IAEA-RGK-1 were mixed in equal weight proportions. The activities calculated using reference values \citep{IAEA} are 1673 Bq$\cdot$kg$^{-1}$ of $^{238}$U, 1083 Bq$\cdot$kg$^{-1}$ of $^{232}$Th and 4669 Bq$\cdot$kg$^{-1}$ of $^{238}$K. This sample was next measured on two systems with correlated and uncorrelated uncertainties, namely, $\mu$Dose and HPGe systems, respectively. The values are reported in Table \ref{tab1}. Note that in the $\mu$Dose system the covariance matrix $\boldsymbol{\Upsigma}$ is known. 

\begin{table*}
\caption{Specific radioactivity measurements using $\mu$Dose and HPGe system from \cite{Tudyka2018a}}
\centering

  \label{tab1}
  \begin{tabular}{ccccc}
    \hline
    ID & Device   & $a_{U-238}$ (Bq$\cdot$kg$^{-1}$) & $a_{Th-232}$ (Bq$\cdot$kg$^{-1}$) & $a_{K-40}$ (Bq$\cdot$kg$^{-1}$)  \\
    \hline
    1 & $\mu$Dose & 1620 $\pm$ 40 & 1100 $\pm$ 60 & 4480 $\pm$  160 \\
    2 & HPGe 	  & 1628 $\pm$ 32 & 1062 $\pm$ 37 & 4610 $\pm$ 110  \\
    3 & known activity & 1673 & 1083 & 4669 \\

    \hline
  \end{tabular}

\end{table*}

\begin{figure*}
\includegraphics[width=185mm]{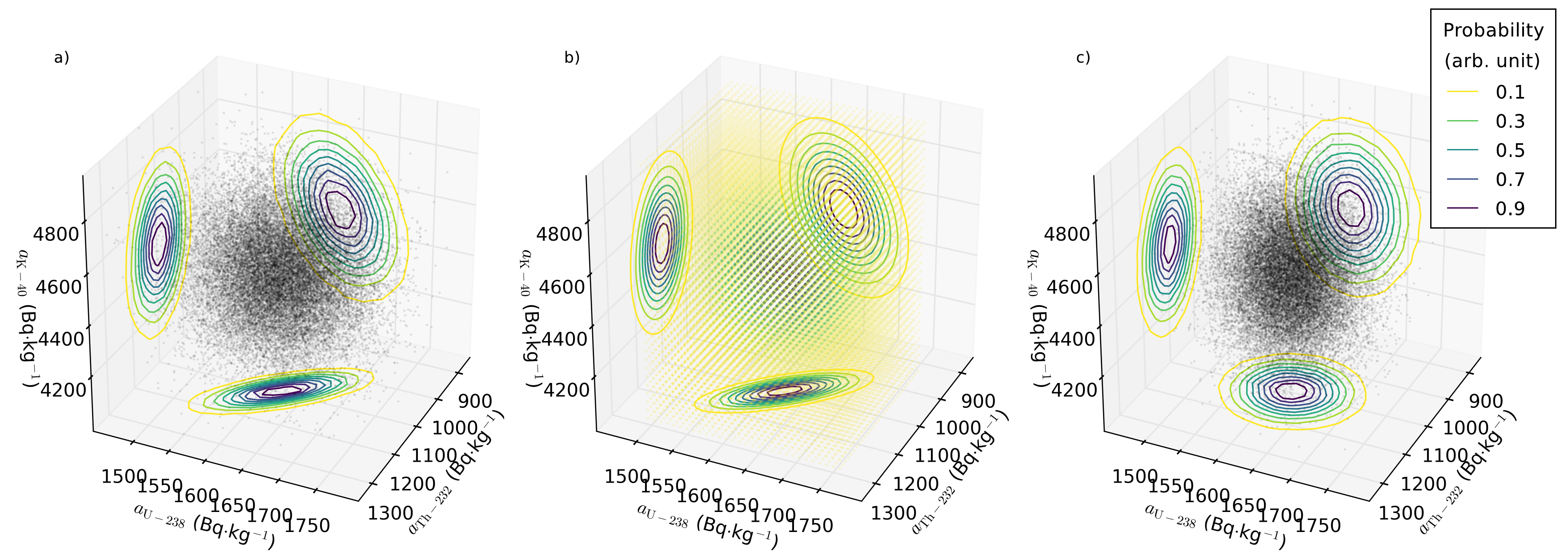}
\caption{\label{Fig2} Three $^{238}$U, $^{232}$Th and $^{40}$K compositions. Isolines are showing projected 2-D PDF contour-plots.
a) A set of 27,000 $^{238}$U, $^{232}$Th and $^{40}$K compositions are drawn by Monte Carlo module from a multivariate normal distribution.  Note that the higher $a_{U-238}$ specific activities correspond to lower $a_{Th-232}$ values and vice versa.
b) A set of 30$\times$30$\times$30 nodes, where the colour of each node indicates probability of each $^{238}$U, $^{232}$Th and $^{40}$K composition. Note that the higher $a_{U-238}$ specific activities correspond to lower $a_{Th-232}$ values and vice versa.
c) A set of 27,000 radioactivities drawn by Monte Carlo module that neglect correlations i.e. setting all off-diagonal elements of $\boldsymbol{\Upsigma}$ to zero. 
}
\end{figure*}

\begin{figure}
\includegraphics[width=85mm]{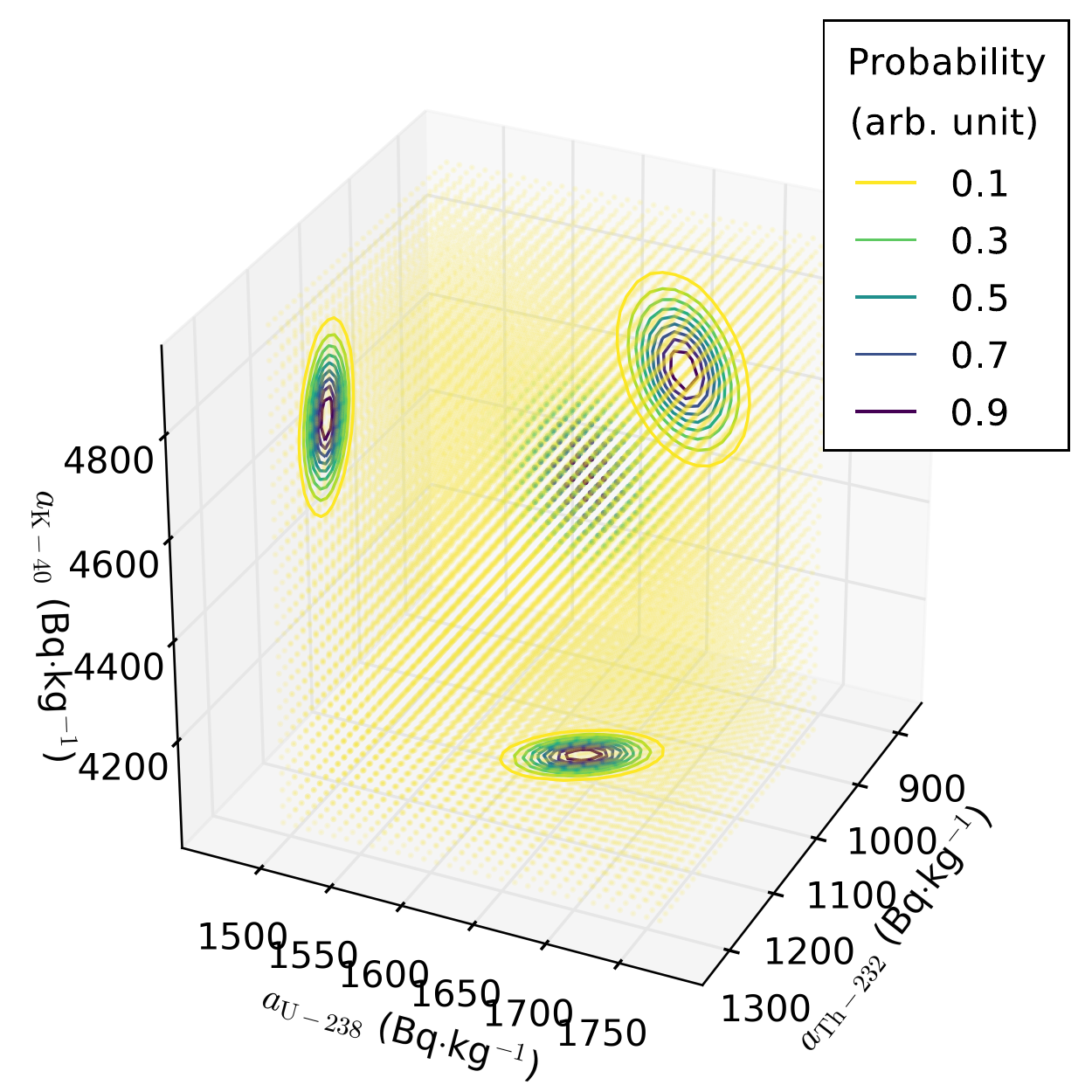}
\caption{\label{Fig3} The probability of $^{238}$U, $^{232}$Th and $^{40}$K compositions for combined data from the $\mu$Dose system and a different HPGe system. Isolines are showing projected 2-D PDF contour-plots.
}
\end{figure}

\indent The first variant includes the Monte Carlo calculations as described in the section \ref{Monte Carlo module} using $\mu$Dose $\boldsymbol{\upmu}$ and $\boldsymbol{\Upsigma}$ data (Table \ref{tab1}). This is illustrated on  Fig. \ref{Fig2}a. The second variant (Fig. \ref{Fig2}b) calculates dose rates and uncertainties as described in the section \ref{Probabilistic module}. We use a set of 30$\times$30$\times$30 nodes where the color of each node indicates the probability. In the third variant (Fig. \ref{Fig2}c) we are drawing $^{238}$U, $^{232}$Th and $^{40}$K values from three independent normal distributions with standard deviations of $\sigma_{U-238}$, $\sigma_{Th-232}$ and $\sigma_{K-40}$. Fig. \ref{Fig2}c shows 27,000 of these points. This variant therefore does not take into consideration the correlation of uncertainties. For these cases (Fig. \ref{Fig2}a-c) we calculate the dose rates and uncertainties using the newest available dose rate conversion factors \citep{Cresswell2018}. The results are summarized in Table \ref{tab2}.\\
\indent Note that in Monte Carlo (Fig. \ref{Fig2}a) and Probabilistic module (Fig. \ref{Fig2}b) the higher $a_{U-238}$ specific activities are lower than the corresponding $a_{Th-232}$ values and vice versa. Hence, the dose rates arising  from the two decay chains will cancel each other out giving lower uncertainties (Table \ref{tab2}, rows 1-2). In the Monte Carlo module, which neglects correlation (Fig. \ref{Fig2}c) no such pattern is observed and the uncertainties are significantly higher (Table \ref{tab2}, row 3). This illustrates the advantage of the proposed approaches in the dose rate calculation that yields better, and lower, error estimates. In this particular measurement, the results shown in row 1 and 2 of Table \ref{tab2} indicate an improvement by a factor of two in the precision, when compared to the uncorrelated values (Table  \ref{tab2}, row 3). \\
\indent The same improvement in dose rate precision arises when we compare measurements performed on $\mu$Dose and HPGe (Table \ref{tab1}). Here in Table \ref{tab1}, $\mu$Dose has higher uncertainties when compared to HPGe. This does not indicate a worse dose rate precision. In fact after dose rate conversion with Monte Carlo (Table  \ref{tab2}, row 1) and the probabilistic module (Table  \ref{tab2}, row 2) we are obtaining higher dose rate precision with $\mu$Dose when compared to HPGe system where uncertainties are uncorrelated (Table  \ref{tab2}, row 4).\\
\indent Finally, in the probabilistic module we can combine data from the $\mu$Dose and a different systems like HPGe. Table  \ref{tab2} row 5 and  (Fig. \ref{Fig3}) shows the merged measurements from Table \ref{tab2}. This increases the precision of the specific radioactivities and the dose rate.

\begin{table*}
\caption{Dose rate precision comparison with three calculation procedures for the IAEA reference material mix RGTh-1, RGU-1 RGK-1 in a 1:1:1 ratio}
\centering
\begin{adjustbox}{width=1\textwidth}

  \label{tab2}
  \begin{tabular}{cccccccccc}
    \hline
    ID & \makecell{Calculation \\ module }  & Device & \makecell{$\dot{D}_\alpha$ \\(Gy$\cdot$ky$^{-1}$)} & $ \displaystyle \frac{u(\dot{D})_\alpha}{\dot{D}_\alpha}$ & \makecell{$\dot{D}_\beta$ \\ (Gy$\cdot$ky$^{-1}$)} &  $ \displaystyle \frac{u(\dot{D})_\beta}{\dot{D}_\beta}$& \makecell{$\dot{D}_\gamma$\\(Gy$\cdot$ky$^{-1}$)}  &  $ \displaystyle \frac{u(\dot{D})_\gamma}{\dot{D}_\gamma}$ & Distribution \\
    \hline

1& Monte Carlo & $\mu$Dose & 568.3$\pm$ 7.1 & 1.2\% & 38.36 $\pm$ 0.55 & 1.4\% & 31.59 $\pm$ 0.46 & 1.5\% & Fig. \ref{Fig2}a\\
2& Probabilistic & $\mu$Dose & 568.1 $\pm$ 7.2 & 1.3\% & 38.36 $\pm$ 0.56 & 1.5\% & 31.56 $\pm$ 0.47 & 1.5\% &Fig. \ref{Fig2}b\\
3& Monte Carlo\textsuperscript{\emph{unc}} & $\mu$Dose & 568 $\pm$ 15 & 2.6\%& 38.36 $\pm$ 0.88 & 2.3\% & 31.59 $\pm$ 0.84 & 2.7\% & Fig. \ref{Fig2}c\\
4& Monte Carlo\textsuperscript{\emph{unc}} & HPGe & 561 $\pm$ 10 & 1.8\% & 38.45 $\pm$ 0.66 & 1.7\% &31.17 $\pm$ 0.56 & 1.8\% &-\\
5& \makecell{Probabilistic\\ with a priori} & \makecell{$\mu$Dose \\and \\HPGe }& 564.1 $\pm$ 6.1 & 1.1\%& 38.53 $\pm$ 0.49 & 1.3\% & 31.40 $\pm$ 0.35 & 1.1\% & Fig. \ref{Fig3}\\

    \hline
  \end{tabular}
\end{adjustbox}

\textsuperscript{\emph{unc}} 27~000 $a_{i, U-238}$ , $a_{i, Th-232}$ and $a_{i, K-40}$ drawn from three independent normal distributions 
\end{table*}

\section{Conclusion}

The built-in Monte Carlo and probabilistic modules for dose rate calculation allow $\alpha$, $\beta$ and $\gamma$ dose rates to be obtained with a higher precision when compared to simplified calculations (Table \ref{tab2}).  Both implemented methods of dose rate calculation in the $\mu$Dose software take into account correlations and are consistent. The Monte Carlo module is somewhat less computer intensive and therefore faster than the probabilistic module. The probabilistic approach on the other hand offers a convenient way of combining measurements from different systems.\\
\indent In the presented example, the dose rate  precision was improved by a factor of two when comparing correlated (Table \ref{tab2}, row 1-2) and uncorrelated (Table \ref{tab2}, row 3) uncertainties assessments. When comparing $\mu$Dose to HPGe, higher uncertainties (Table \ref{tab1}) do not indicate worse dose rate precision. In fact, after dose rate conversion with Monte Carlo (Table  \ref{tab2}, row 1) and the probabilistic module (Table  \ref{tab2} row 2) we are obtaining a higher dose rate precision with $\mu$Dose when compared to the HPGe system, where uncertainties are uncorrelated (Table  \ref{tab2}, row 4). This also means that to obtain the same dose rate uncertainty counting time can be significantly reduced. \\
\indent The Bayesian module can further help to increase the precision if results of independent measurements are available (Table  \ref{tab2}, row 5).\\
\indent The inbuilt dose rate module allows for accounting of the $\alpha$ particle effectiveness with $a$-value, relative grain sizes, chemical etching, water content, relative sample position in the investigated profile and cosmic dose rate. These corrections are needed to obtain reliable annual dose rates in trapped charge dating.\\
\indent Finally, the $\mu$Dose dose rate module is fully controlled by an intuitive graphical interface. Moreover, for the user convenience, the input data are pre-filled with the most recently obtained values. This greatly simplifies the dose rate calculation procedure.

\section{Acknowledgements}

The development of the pulse analyzer used in the  $\mu$Dose system  was supported with the LIDER/001/404/L-4/2013 grant given by the Polish National Centre for Research and Development.
Currently the project is co-financed by the Ministry of Science and Higher Education from ''Incubator of Innovation+'' programme within the framework of the Smart Growth Operational Programme, Action 4.4 Potential increase of human resources of the R\&D sector.

\section{References}

\bibliography{dose_rate_calculator}

\end{document}